\documentclass[aps,prb,twocolumn,groupedaddress,showpacs,superscriptaddress]{revtex4-1}

\usepackage{graphicx}
\usepackage{array}
\usepackage{multirow}
\usepackage{amsmath}
\usepackage{amssymb}
\usepackage{graphics}
\usepackage{color}

\begin{document}


\title{Lattice dynamics of KNi$_{2}$Se$_2$}

\author{N. Lazarevi\'c}
 \affiliation{Center for Solid State Physics and New Materials, Institute of Physics Belgrade, University of Belgrade, Pregrevica 118, 11080 Belgrade, Serbia}

\author{M. Radonji\'c}
\affiliation{Scientific Computing Laboratory, Institute of Physics Belgrade, University of Belgrade, Pregrevica 118, 11080 Belgrade, Serbia}

\author{M. \v{S}\'cepanovi\'c}
 \affiliation{Center for Solid State Physics and New Materials, Institute of Physics Belgrade, University of Belgrade, Pregrevica 118, 11080 Belgrade, Serbia}

\author{Hechang Lei}
\affiliation{Condensed Matter Physics and Materials Science Department, Brookhaven
National Laboratory, Upton, New York 11973-5000, USA}

\author{D. Tanaskovi\'c}
\affiliation{Scientific Computing Laboratory, Institute of Physics Belgrade, University of Belgrade, Pregrevica 118, 11080 Belgrade, Serbia}

\author{C. Petrovic}
\affiliation{Condensed Matter Physics and Materials Science Department, Brookhaven
National Laboratory, Upton, New York 11973-5000, USA}

\author{Z. V. Popovi\'c}
\affiliation{Center for Solid State Physics and New Materials, Institute of Physics Belgrade, University of Belgrade,
Pregrevica 118, 11080 Belgrade, Serbia}

\date{\today}

\begin{abstract}
We report the first principle calculations of the lattice dynamics of KNi$_{2}$Se$_2$ together with the Raman scattering study.\ We have observed three out of four Raman active modes predicted by the factor group analysis. Calculated phonon frequencies are in
good agreement with experimental findings.\ Contrary to its iron counterpart (K$_x$Fe$_{2-y}$Se$_2$), K$_{0.95}$Ni$_{1.86}$Se$_2$ does not show vacancy ordering.\
\end{abstract}

\maketitle

\section{Introduction}

The discovery of superconductivity in the iron materials has aroused great interest among researches to study the physical properties of these
materials which are dominated by the layers of Fe atoms surrounded by the elements of pnictogen (As,P) or the chalcogen group (Se,Te).\cite{Kamihara,Rotter,Kuo-Wei,Fong-Chi,Wang} The recently
discovered superconductivity in the alkali-doped iron selenide layered compounds with $T_c\sim$ 33 K brings forth some unique characteristics that are absent in
the other iron-based superconductors.\cite{Guo}\ These include the presence of the iron vacancies and their ordering, the antiferromagnetically ordered insulating phases and a very high N\'{e}el transition temperature.\cite{Bao,Zhang,Ye,Ryan,Shermadini}

Nickel pnictides have recently attracted a lot of attention,\cite{Neilson} despite of the low
critical superconducting temperature, much lower than in iron based pnictides. The cause of such significant distinction in $T_c$ value is not clear. It could be
the consequence of the different superconducting mechanisms, or different values of the material parameters responsible for superconductivity. Typically, these
materials display a very reach phase diagram including phases with magnetic ordering, or heavy fermion phase, which is typically accompanied by a superconducting phase
at low temperatures. KNi$_{2}$Se$_2$ shows a putative local charge density wave (LCDW) state which persists up to 300 K, followed by the magnetic field independent heavy fermion
phase below 40 K and a superconducting phase below $T_c$=0.8 K.\cite{Neilson} The superconducting phase is very sensitive on stoichiometry.\cite{HLei}\ Even small deficiency
of K and Ni atoms leads to the absence of superconducting phase down to $0.3$ K.\ Therefore, in order to understand the low temperature transport and thermodynamic properties of this material, a full knowledge of the lattice dynamics is necessary.\ To the best of our knowledge phonon properties of this compound are unknown.

In this paper we address the lattice dynamics of KNi$_{2}$Se$_2$.\
The first principles lattice dynamics calculations were performed within density functional perturbation theory\cite{DFPT} (DFPT) using QUANTUM ESPRESSO\cite{QE} package.\
The polarized Raman scattering measurements were performed in a wide temperature range.\
Three out of four Raman active modes predicted by the symmetry considerations are observed and assigned.\

\section{Experiment and numerical calculations}

Single crystal growth and characterization of the K$_{0.95}$Ni$_{1.86}$Se$_2$ samples were described in previous report.\cite{HLei}\ Raman scattering measurements were performed on freshly cleaved samples using a JY T64000 and TriVista 557 Raman systems in backscattering micro-Raman configuration. The 514.5 nm line of a mixed Ar$^+$/Kr$^+$ gas laser was used as an excitation source.\ The corresponding excitation power density was less than 0.2 kW/cm$^2$.\ Low temperature measurements were performed using KONTI CryoVac continuous flow cryostat with 0.5 mm thick window.\ Measurements in the optical phonon region of
K$_{0.95}$Ni$_{1.86}$Se$_2$ (30 - 350 cm$^{-1}$) were performed using 1800/1800/1800 groves/mm gratings configuration of JY T64000 system and 900/900/2400 groves/mm gratings configuration of TriVista 557 system.

\begin{figure}
\includegraphics[width=0.4\textwidth]{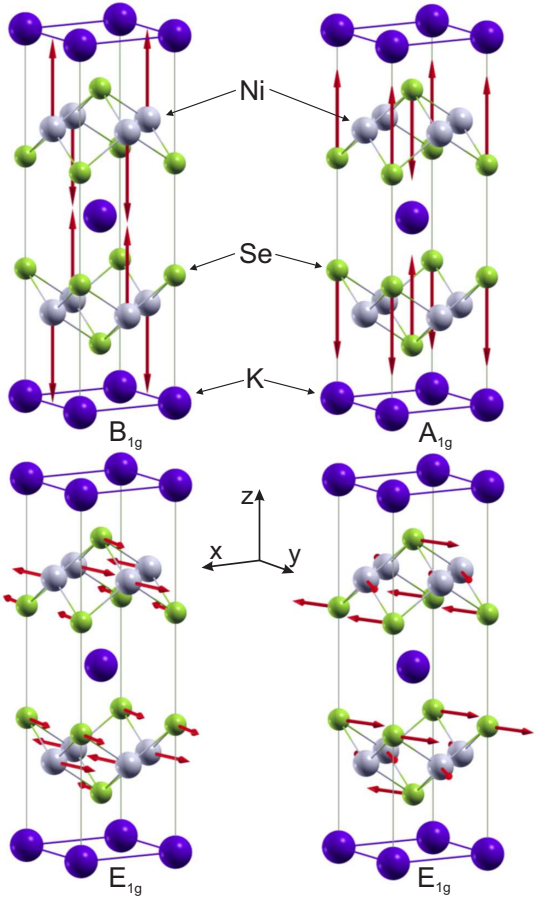}
\caption{(Color online) Displacement patterns of the Raman-active vibrational
modes of KNi$_2$Se$_2$.\ }
\label{vib}
\end{figure}

We have performed calculations of the lattice dynamics of the KNi$_2$Se$_2$ within the DFPT\cite{DFPT} using QUANTUM ESPRESSO\cite{QE} package.\ K$_{}$Ni$_{2}$Se$_2$ crystallizes in the tetragonal ThCr$_2$Si$_2$-type of crystal structure (I4/mmm space group with the unit cell parameters $a=3.9089(8)$ \AA, $c=13.4141(5)$ \AA, $z=0.35429(2)$).\cite{Neilson,HLei}\  Potassium atoms are at $2a: (0,0,0)$, Ni atoms at $4d: (0,\frac{1}{2},\frac{1}{4})$, and Se atoms at $4e: (0,0,z)$ Wyckoff positions.\
In our calculations we have used the ultra-soft Projector Augmented-Wave (PAW) pseudo-potentials calculated with Perdew-Burke-Ernzerhof (PBE) exchange-correlation
functional and nonlinear core correction.\ We have carried out the relaxation of the structural parameters until all forces acting on the individual atom in the
unit cell became smaller than $5\times 10^{-6}$ Ry/a.u. and all the stresses to the unit cell were smaller than $0.01$ kbar.\ The relaxed structural parameters are $a=3.9490$ \AA, $c=13.0552$ \AA, $z=0.35250$ and they are in good agreement (within few precent) with the experimentally measured values.\ The electronic calculations are
performed on $16\times 16 \times 16$ Monkhorst-Pack {\bf k}-space mesh, with the kinetic-energy cutoff of $41$ Ry and the charge-density cutoff of $236$ Ry and the gaussian
smearing $0.005$.\ The obtained $\Gamma$ point phonon energies are listed in Table~\ref{tab.1}.\
The normal modes of all four Raman-active phonons are shown in Figure~\ref{vib}.\ As can be seen from Figure~\ref{vib} the A$_{1g}$ (B$_{1g}$) mode represents the vibrations of the Se (Ni) ions along the $c$-axis, whereas the E$_g$ modes involve the vibration both Ni and Se ions within the $ab$ plane.

\section{Results and discussion}

\begin{figure}
\includegraphics[width=0.4\textwidth]{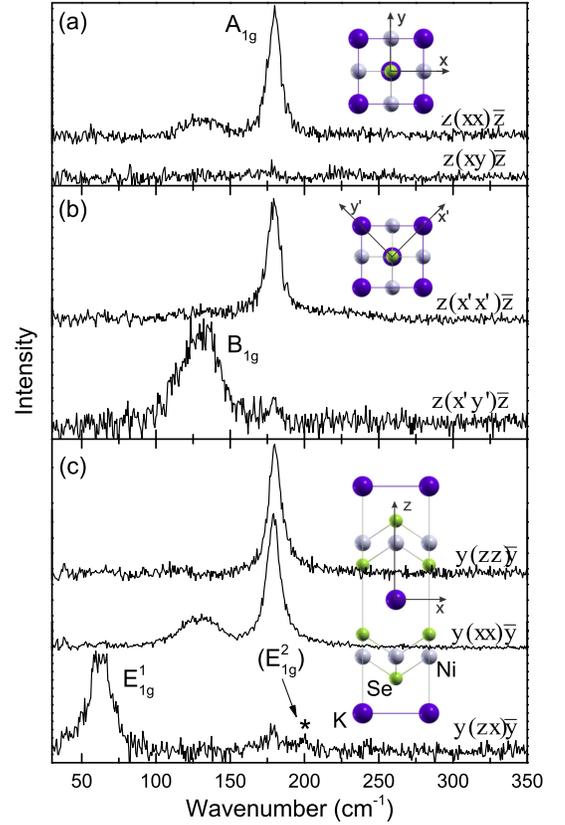}
\caption{(Color online) Raman scattering spectra of K$_{0.95}$Ni$_{1.86}$Se$_2$ single crystals measured at room temperature using JY T64000 Raman system in various scattering configurations ($\textbf{x}=[100]$, $\textbf{y}=[010]$, $\textbf{x}'=1/\sqrt{2}[110]$, $\textbf{y}'=1/\sqrt{2}[1\bar{1}0]$, $\textbf{z}=[001]$).\ }
\label{fig1}
\end{figure}


\begin{table*}[t]\footnotesize
\caption{Top panel gives the type of atoms together with their Wyckoff's positions, each site contributions to the $\Gamma$ point phonons, as well as Raman tensors, phonon activities and selection rules for KNi$_2$Se$_2$ (I4/mmm space group).\ Lower panel of the table lists experimental (at room temperature) and calculated phonon mode frequencies and their activity.}
\label{tab.1}
\begin{ruledtabular}
\centering
\begin{tabular}{ccccc}
Atoms  & Wyckoff position & \multicolumn{3}{c}{Irreducible representations}                             \\ \hline
K      & 2a               & \multicolumn{3}{c}{A$_{2u}$+E$_u$ }                                         \\
Ni     & 4d               & \multicolumn{3}{c}{A$_{2u}$+B$_{1g}$+E$_g$+E$_u$}                           \\
Se     & 4e               & \multicolumn{3}{c}{A$_{1g}$+A$_{2u}$+E$_g$+E$_u$}                           \\
\multicolumn{5}{c}{Raman tensors}\\

$\hat{R}_{A_{1g}}$=$\left(\begin{matrix}
a & 0 & 0 \\
0 & a & 0 \\
0 & 0 & b%
\end{matrix}%
\right)$ &
$\hat{R}_{B_{1g}}$=$\left(\begin{matrix}
c & 0 & 0 \\
0 & -c & 0 \\
0 & 0  & 0%
\end{matrix}%
\right)$ &
\multicolumn{3}{c}{$\hat{R}_{E_{g}}$=$\left(\begin{matrix}
0 & 0 & e \\
0 & 0 & 0 \\
e & 0 & 0%
\end{matrix}%
\right)$  \hspace{0.7cm}    $\hat{R}_{E_{g}}$=$\left(\begin{matrix}
0 & 0 & 0 \\
0 & 0 & f \\
0 & f & 0%
\end{matrix}%
\right)$}    \\
\multicolumn{5}{c}{Activity and selection rules}\\
\multicolumn{5}{c}{$\Gamma_{Raman}$=A$_{1g}$($\alpha_{xx+yy}$, $\alpha_{zz}$)+B$_{1g}$($\alpha_{xx-yy}$)+2E$_g$($\alpha_{xz}$, $\alpha_{yz}$)} \\  \multicolumn{5}{c}{$\Gamma_{infrared}$=2A$_{2u}$(\textbf{E}$\parallel$\textbf{z})+2E$_u$(\textbf{E}$\parallel$\textbf{x},\textbf{E}$\parallel$\textbf{y})}\\
\multicolumn{5}{c}{$\Gamma_{acoustic}$=A$_u$+E$_u$} \\ \hline

Symmetry & Activity & Experiment (cm$^{-1}$) & Calculations (cm$^{-1}$) & Main atomic displacements \\ \hline
A$_{1g}$  & Raman    &     179    &     189.4                        & Se (z)                       \\
B$_{1g}$  & Raman    &     134    &     133.8                        & Ni (z)                       \\
E$_{g}^1$   & Raman    &     63     &     35.4                         & Ni(xy), Se(xy)               \\
E$_{g}^2$   & Raman    &     (201)  &     203.9                        & Ni(xy), Se(xy)               \\
A$_{2u}^1$  & IR       &            &     116.4                        & K(z), Se(-z)                 \\
A$_{2u}^2$  & IR       &            &     220.8                        & Ni(z), K(-z)                 \\
E$_{u}^1$   & IR       &            &     105.1                        & K(xy)                        \\
E$_{u}^2$   & IR       &            &     208.3                        & Ni(xy), Se(-xy)              \\
\end{tabular}
\end{ruledtabular}
\end{table*}

Symmetry considerations predict four Raman-active phonons: A$_{1g}$, B$_{1g}$ and 2E$_{g}$ (Table~\ref{tab.1}) for the KNi$_{2}$Se$_2$.\ However, ordering of the vacancies may reduce the symmetry to I4/m.\ This results in an increase of the number of Raman active modes as it was shown for the K$_{x}$Fe$_{2-y}$Se$_2$\cite{Lazarevic1} and K$_{x}$Fe$_{2-y}$S$_2$\cite{Lazarevic2}.\
Figure~\ref{fig1} shows room temperature polarized Raman sca\-ttering spectra of K$_{0.95}$Ni$_{1.86}$Se$_2$ single crystals.\
Only three Raman active modes are observed in the Raman spectra for different sample orientations.\ This finding supports the high symmetry (I4/mmm space group) of the K$_{0.95}$Ni$_{1.86}$Se$_2$ structure without the Ni vacancy ordering as opposed to the K$_{x}$Fe$_{2-y}$Se$_2$ case.\cite{Lazarevic1}

According to the selection rules, the Raman scattering spectra measured from the $ab$ plane of the sample may contain only A$_{1g}$ and B$_{1g}$ modes.\
The A$_{1g}$ mode can be observed for any orientation of the incident light polarization $\textbf{e}_i$ provided that the scattered light polarization $\textbf{e}_s$ is parallel to it ($\textbf{e}_s\parallel \textbf{e}_i$) and will vanish in any crossed polarization configuration ($\textbf{e}_s\perp \textbf{e}_i$).\ On the other hand, the intensity of the B$_{1g}$ mode strongly depends on the sample orientation
($I_{B_{1g}}(\Theta)\sim|c|^2\cos^2(\Theta+2\beta)$ where $\Theta=\angle(\textbf{e}_i,\textbf{e}_s)$ and $\beta=\angle(\textbf{e}_i,\textbf{x})$)\cite{Lazarevic1}.\
When the sample is oriented so that $\textbf{e}_i\parallel \textbf{x}$, (see Figure~\ref{fig1}~(a)) one can expect the appearance of both the A$_{1g}$ and B$_{1g}$ modes in the parallel and their
absence for a cross polarization.\ In order to separate the A$_{1g}$ from the B$_{1g}$ symmetry mode, incident light polarization should be parallel to $\textbf{x}'=1/\sqrt{2}[110]$ axis of the crystal (see Figure~\ref{fig1}~(b)).\ The Raman mode at about 179 cm$^{-1}$ has been observed in the
parallel, but not in the cross polarization configuration and, consequently it is assigned as the A$_{1g}$ mode.\ The mode at about 134 cm$^{-1}$ has been observed in the cross
but not in the parallel polarization configuration and consequently is assigned as the B$_{1g}$ mode.\

Observation of the E$_g$ symmetry modes, in the case of the tetragonal crystal symmetry, requires performing measurements in the $ac$ plane of the sample.
According to the selection rules, for the parallel polarization configuration with $\textbf{e}_i\parallel \textbf{z}$ the A$_{1g}$ mode appearance is the only one to be expected, whereas both the A$_{1g}$ and B$_{1g}$ modes are expected to be observable in the case of $\textbf{e}_i\parallel \textbf{x}$ (see Figure~\ref{fig1}~(c)).\ In the cross polarization configuration only the E$_g$ modes can be observed.
Consequently, mode at around 63 cm$^{-1}$ (see Figure~\ref{fig1}~(c)) has been assigned as E$_g^1$ symmetry one.\ In addition, weak peak-like feature has been observed at around 201 cm$^{-1}$ (denoted by asterisk in Figure~\ref{fig1}~(c)).\ However, assignation of this feature cannot be unambiguously performed because of the extremely low intensity, although it falls in the region where the appearance of the E$_g^2$ mode is expected (see Table~\ref{tab.1}).\ The frequencies of the observed modes are in good agree\-ment with our calculations (see Table~\ref{tab.1}).\

\begin{figure}
\includegraphics[width=0.4\textwidth]{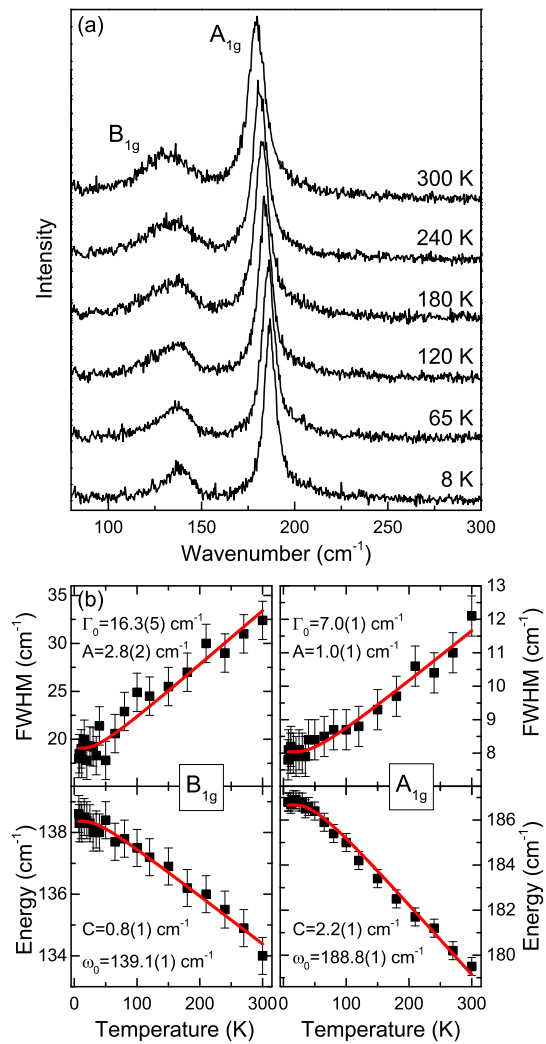}
\caption{(Color online)(a) Temperature-dependent unpolarized Raman scattering
spectra of the K$_{0.95}$Ni$_{1.86}$Se$_2$ single crystals measured from the $ab$ plane of the sample using TriVista 557 Raman system.\ (b) Energy and linewidth of A$_{1g}$
and B$_{1g}$ modes as a function of temperature.\ Solid lines show the expected behavior due to anharmonic phonon decay (see the text).}
\label{fig2}
\end{figure}

Figure.~\ref{fig2}~(a) shows unpolarized Raman spectra of K$_{0.95}$Ni$_{1.86}$Se$_2$ single crystal, measured from the
$ab$ plane of the sample at various temperatures.\
No observable change has been observed in the spectra near the local CDW to heavy fermion transition temperature ($T\sim 40$ K).\
The A$_{1g}$ and B$_{1g}$ symmetry modes energy and full-width at half maximum (FWHM) temperature dependance are presented in Figure~\ref{fig2}~(b).

Temperature dependance of the phonon mode energy, $\Omega(T)$,  and linewidth, $\Gamma(T)$, are usually governed by phonon-phonon interaction (anharmonic effects).\ For simplicity we assume a
symmetric decay of the low lying optical phonon into two acoustic phonons:\cite{anh1}
\begin{equation}
\Omega (T)=\Omega_0-C\bigg(1+\frac{2}{e^{x}-1}\bigg)
\label{eq1}
\end{equation}
\begin{equation}
\Gamma (T)=\Gamma_0+A\bigg(1+\frac{2}{e^{x}-1}\bigg),
\label{eq2}
\end{equation}
where $\Omega_0$ is the Raman mode energy, $A$ and $C$ are the anharmonic constants, $x=\hbar\Omega_0/2k_BT$.\
In the case of semiconducting and insulating materials, $A$ is usually the only parameter needed for describing the temperature dependence of the linewidth.\
Phonons may also couple with other elementary excitations i.e. electrons, in which case additional term $\Gamma_0$ must be included.\
$\Gamma_0$ term also includes the contributions from scattering on defects.\

Red lines in Figure~\ref{fig2}~(b) represent calculated spectra by using Eqs.~(\ref{eq1}) and (\ref{eq2}).\
Although there is a good agree\-ment with the experimental data, large value of $\Gamma_0$ parameter, especially for the B$_{1g}$ phonon, together with the clear asymmetry of this mode, points out to the possible contribution from the interaction of the phonons with some other excitations (i.e. electrons).\cite{LazarevicFano}\ However, a clear non-stoichiometry of the studied single crystals indicates that the origin of the increased width and asymmetry of the B$_{1g}$ mode is more likely due to disorder that breaks the conservation of the momentum during the Raman scattering process enabling contributions of finite wavevector phonons to the Raman spectra.

\section{Conclusion}

We have performed the Raman scattering study and the lattice dynamics calculations of the K$_{}$Ni$_{2}$Se$_2$.\
By analyzing polarized Raman scattering spectra of K$_{0.95}$Ni$_{1.86}$Se$_2$ single crystals, we have identified three out of four Raman active modes predicted by the factor group analysis.\
Frequencies of these modes are in good agreement with the lattice dynamics results.\
Contrary to its counterpart K$_{x}$Fe$_{2-y}$Se$_2$, K$_{0.95}$Ni$_{1.86}$Se$_2$ did not show Ni vacancy ordering.\
Temperature dependent study revealed no significant changes in the Raman spectra near the local CDW to heavy fermion phase transition temperature.\

\section*{Acknowledgment}

This work was supported by the Serbian Ministry of Education, Science and Technological Development under Projects ON171032, III45018 and ON171017. Part of this work was carried out at the Brookhaven
National Laboratory which is operated for the Office of Basic Energy
Sciences, U.S. Department of Energy by Brookhaven Science Associates
(DE-Ac02-98CH10886). Numerical simulations were run on the AEGIS
e-Infrastructure, supported in part by FP7 projects EGI-InSPIRE,
PRACE-1IP and HP-SEE.


\begin{thebibliography}{19}%
\makeatletter
\providecommand \@ifxundefined [1]{%
 \@ifx{#1\undefined}
}%
\providecommand \@ifnum [1]{%
 \ifnum #1\expandafter \@firstoftwo
 \else \expandafter \@secondoftwo
 \fi
}%
\providecommand \@ifx [1]{%
 \ifx #1\expandafter \@firstoftwo
 \else \expandafter \@secondoftwo
 \fi
}%
\providecommand \natexlab [1]{#1}%
\providecommand \enquote  [1]{``#1''}%
\providecommand \bibnamefont  [1]{#1}%
\providecommand \bibfnamefont [1]{#1}%
\providecommand \citenamefont [1]{#1}%
\providecommand \href@noop [0]{\@secondoftwo}%
\providecommand \href [0]{\begingroup \@sanitize@url \@href}%
\providecommand \@href[1]{\@@startlink{#1}\@@href}%
\providecommand \@@href[1]{\endgroup#1\@@endlink}%
\providecommand \@sanitize@url [0]{\catcode `\\12\catcode `\$12\catcode
  `\&12\catcode `\#12\catcode `\^12\catcode `\_12\catcode `\%12\relax}%
\providecommand \@@startlink[1]{}%
\providecommand \@@endlink[0]{}%
\providecommand \url  [0]{\begingroup\@sanitize@url \@url }%
\providecommand \@url [1]{\endgroup\@href {#1}{\urlprefix }}%
\providecommand \urlprefix  [0]{URL }%
\providecommand \Eprint [0]{\href }%
\providecommand \doibase [0]{http://dx.doi.org/}%
\providecommand \selectlanguage [0]{\@gobble}%
\providecommand \bibinfo  [0]{\@secondoftwo}%
\providecommand \bibfield  [0]{\@secondoftwo}%
\providecommand \translation [1]{[#1]}%
\providecommand \BibitemOpen [0]{}%
\providecommand \bibitemStop [0]{}%
\providecommand \bibitemNoStop [0]{.\EOS\space}%
\providecommand \EOS [0]{\spacefactor3000\relax}%
\providecommand \BibitemShut  [1]{\csname bibitem#1\endcsname}%
\let\auto@bib@innerbib\@empty
\bibitem [{\citenamefont {Kamihara}\ \emph {et~al.}(2008)\citenamefont
  {Kamihara}, \citenamefont {Watanabe}, \citenamefont {Hirano},\ and\
  \citenamefont {Hosono}}]{Kamihara}%
  \BibitemOpen
  \bibfield  {author} {\bibinfo {author} {\bibfnamefont {Y.}~\bibnamefont
  {Kamihara}}, \bibinfo {author} {\bibfnamefont {T.}~\bibnamefont {Watanabe}},
  \bibinfo {author} {\bibfnamefont {M.}~\bibnamefont {Hirano}}, \ and\ \bibinfo
  {author} {\bibfnamefont {H.}~\bibnamefont {Hosono}},\ }\href {\doibase
  10.1021/ja800073m} {\bibfield  {journal} {\bibinfo  {journal} {Journal of the
  American Chemical Society}\ }\textbf {\bibinfo {volume} {130}},\ \bibinfo
  {pages} {3296} (\bibinfo {year} {2008})},\ \Eprint
  {http://arxiv.org/abs/http://pubs.acs.org/doi/pdf/10.1021/ja800073m}
  {http://pubs.acs.org/doi/pdf/10.1021/ja800073m} \BibitemShut {NoStop}%
\bibitem [{\citenamefont {Rotter}\ \emph {et~al.}(2008)\citenamefont {Rotter},
  \citenamefont {Tegel},\ and\ \citenamefont {Johrendt}}]{Rotter}%
  \BibitemOpen
  \bibfield  {author} {\bibinfo {author} {\bibfnamefont {M.}~\bibnamefont
  {Rotter}}, \bibinfo {author} {\bibfnamefont {M.}~\bibnamefont {Tegel}}, \
  and\ \bibinfo {author} {\bibfnamefont {D.}~\bibnamefont {Johrendt}},\ }\href
  {\doibase 10.1103/PhysRevLett.101.107006} {\bibfield  {journal} {\bibinfo
  {journal} {Phys. Rev. Lett.}\ }\textbf {\bibinfo {volume} {101}},\ \bibinfo
  {pages} {107006} (\bibinfo {year} {2008})}\BibitemShut {NoStop}%
\bibitem [{\citenamefont {Yeh}\ \emph {et~al.}(2008)\citenamefont {Yeh},
  \citenamefont {Huang}, \citenamefont {lin Huang}, \citenamefont {Chen},
  \citenamefont {Hsu}, \citenamefont {Wu}, \citenamefont {Lee}, \citenamefont
  {Chu}, \citenamefont {Chen}, \citenamefont {Luo}, \citenamefont {Yan},\ and\
  \citenamefont {Wu}}]{Kuo-Wei}%
  \BibitemOpen
  \bibfield  {author} {\bibinfo {author} {\bibfnamefont {K.-W.}\ \bibnamefont
  {Yeh}}, \bibinfo {author} {\bibfnamefont {T.-W.}\ \bibnamefont {Huang}},
  \bibinfo {author} {\bibfnamefont {Y.}~\bibnamefont {lin Huang}}, \bibinfo
  {author} {\bibfnamefont {T.-K.}\ \bibnamefont {Chen}}, \bibinfo {author}
  {\bibfnamefont {F.-C.}\ \bibnamefont {Hsu}}, \bibinfo {author} {\bibfnamefont
  {P.~M.}\ \bibnamefont {Wu}}, \bibinfo {author} {\bibfnamefont {Y.-C.}\
  \bibnamefont {Lee}}, \bibinfo {author} {\bibfnamefont {Y.-Y.}\ \bibnamefont
  {Chu}}, \bibinfo {author} {\bibfnamefont {C.-L.}\ \bibnamefont {Chen}},
  \bibinfo {author} {\bibfnamefont {J.-Y.}\ \bibnamefont {Luo}}, \bibinfo
  {author} {\bibfnamefont {D.-C.}\ \bibnamefont {Yan}}, \ and\ \bibinfo
  {author} {\bibfnamefont {M.-K.}\ \bibnamefont {Wu}},\ }\href
  {http://stacks.iop.org/0295-5075/84/i=3/a=37002} {\bibfield  {journal}
  {\bibinfo  {journal} {EPL (Europhysics Letters)}\ }\textbf {\bibinfo {volume}
  {84}},\ \bibinfo {pages} {37002} (\bibinfo {year} {2008})}\BibitemShut
  {NoStop}%
\bibitem [{\citenamefont {Hsu}\ \emph {et~al.}(2008)\citenamefont {Hsu},
  \citenamefont {Luo}, \citenamefont {Yeh}, \citenamefont {Chen}, \citenamefont
  {Huang}, \citenamefont {Wu}, \citenamefont {Lee}, \citenamefont {Huang},
  \citenamefont {Chu}, \citenamefont {Yan},\ and\ \citenamefont
  {Wu}}]{Fong-Chi}%
  \BibitemOpen
  \bibfield  {author} {\bibinfo {author} {\bibfnamefont {F.-C.}\ \bibnamefont
  {Hsu}}, \bibinfo {author} {\bibfnamefont {J.-Y.}\ \bibnamefont {Luo}},
  \bibinfo {author} {\bibfnamefont {K.-W.}\ \bibnamefont {Yeh}}, \bibinfo
  {author} {\bibfnamefont {T.-K.}\ \bibnamefont {Chen}}, \bibinfo {author}
  {\bibfnamefont {T.-W.}\ \bibnamefont {Huang}}, \bibinfo {author}
  {\bibfnamefont {P.~M.}\ \bibnamefont {Wu}}, \bibinfo {author} {\bibfnamefont
  {Y.-C.}\ \bibnamefont {Lee}}, \bibinfo {author} {\bibfnamefont {Y.-L.}\
  \bibnamefont {Huang}}, \bibinfo {author} {\bibfnamefont {Y.-Y.}\ \bibnamefont
  {Chu}}, \bibinfo {author} {\bibfnamefont {D.-C.}\ \bibnamefont {Yan}}, \ and\
  \bibinfo {author} {\bibfnamefont {M.-K.}\ \bibnamefont {Wu}},\ }\href
  {\doibase 10.1073/pnas.0807325105} {\bibfield  {journal} {\bibinfo  {journal}
  {Proceedings of the National Academy of Sciences}\ }\textbf {\bibinfo
  {volume} {105}},\ \bibinfo {pages} {14262} (\bibinfo {year} {2008})},\
  \Eprint
  {http://arxiv.org/abs/http://www.pnas.org/content/105/38/14262.full.pdf+html}
  {http://www.pnas.org/content/105/38/14262.full.pdf+html} \BibitemShut
  {NoStop}%
\bibitem [{\citenamefont {Wang}\ \emph {et~al.}(2008)\citenamefont {Wang},
  \citenamefont {Liu}, \citenamefont {Lv}, \citenamefont {Gao}, \citenamefont
  {Yang}, \citenamefont {Yu}, \citenamefont {Li},\ and\ \citenamefont
  {Jin}}]{Wang}%
  \BibitemOpen
  \bibfield  {author} {\bibinfo {author} {\bibfnamefont {X.}~\bibnamefont
  {Wang}}, \bibinfo {author} {\bibfnamefont {Q.}~\bibnamefont {Liu}}, \bibinfo
  {author} {\bibfnamefont {Y.}~\bibnamefont {Lv}}, \bibinfo {author}
  {\bibfnamefont {W.}~\bibnamefont {Gao}}, \bibinfo {author} {\bibfnamefont
  {L.}~\bibnamefont {Yang}}, \bibinfo {author} {\bibfnamefont {R.}~\bibnamefont
  {Yu}}, \bibinfo {author} {\bibfnamefont {F.}~\bibnamefont {Li}}, \ and\
  \bibinfo {author} {\bibfnamefont {C.}~\bibnamefont {Jin}},\ }\href {\doibase
  10.1016/j.ssc.2008.09.057} {\bibfield  {journal} {\bibinfo  {journal} {Solid
  State Communications}\ }\textbf {\bibinfo {volume} {148}},\ \bibinfo {pages}
  {538 } (\bibinfo {year} {2008})}\BibitemShut {NoStop}%
\bibitem [{\citenamefont {Guo}\ \emph {et~al.}(2010)\citenamefont {Guo},
  \citenamefont {Jin}, \citenamefont {Wang}, \citenamefont {Wang},
  \citenamefont {Zhu}, \citenamefont {Zhou}, \citenamefont {He},\ and\
  \citenamefont {Chen}}]{Guo}%
  \BibitemOpen
  \bibfield  {author} {\bibinfo {author} {\bibfnamefont {J.}~\bibnamefont
  {Guo}}, \bibinfo {author} {\bibfnamefont {S.}~\bibnamefont {Jin}}, \bibinfo
  {author} {\bibfnamefont {G.}~\bibnamefont {Wang}}, \bibinfo {author}
  {\bibfnamefont {S.}~\bibnamefont {Wang}}, \bibinfo {author} {\bibfnamefont
  {K.}~\bibnamefont {Zhu}}, \bibinfo {author} {\bibfnamefont {T.}~\bibnamefont
  {Zhou}}, \bibinfo {author} {\bibfnamefont {M.}~\bibnamefont {He}}, \ and\
  \bibinfo {author} {\bibfnamefont {X.}~\bibnamefont {Chen}},\ }\href {\doibase
  10.1103/PhysRevB.82.180520} {\bibfield  {journal} {\bibinfo  {journal} {Phys.
  Rev. B}\ }\textbf {\bibinfo {volume} {82}},\ \bibinfo {pages} {180520}
  (\bibinfo {year} {2010})}\BibitemShut {NoStop}%
\bibitem [{\citenamefont {Wei}\ \emph {et~al.}(2011)\citenamefont {Wei},
  \citenamefont {Qing-Zhen}, \citenamefont {Gen-Fu}, \citenamefont {Green},
  \citenamefont {Du-Ming}, \citenamefont {Jun-Bao},\ and\ \citenamefont
  {Yi-Ming}}]{Bao}%
  \BibitemOpen
  \bibfield  {author} {\bibinfo {author} {\bibfnamefont {B.}~\bibnamefont
  {Wei}}, \bibinfo {author} {\bibfnamefont {H.}~\bibnamefont {Qing-Zhen}},
  \bibinfo {author} {\bibfnamefont {C.}~\bibnamefont {Gen-Fu}}, \bibinfo
  {author} {\bibfnamefont {M.~A.}\ \bibnamefont {Green}}, \bibinfo {author}
  {\bibfnamefont {W.}~\bibnamefont {Du-Ming}}, \bibinfo {author} {\bibfnamefont
  {H.}~\bibnamefont {Jun-Bao}}, \ and\ \bibinfo {author} {\bibfnamefont
  {Q.}~\bibnamefont {Yi-Ming}},\ }\href
  {http://stacks.iop.org/0256-307X/28/i=8/a=086104} {\bibfield  {journal}
  {\bibinfo  {journal} {Chinese Physics Letters}\ }\textbf {\bibinfo {volume}
  {28}},\ \bibinfo {pages} {086104} (\bibinfo {year} {2011})}\BibitemShut
  {NoStop}%
\bibitem [{\citenamefont {Zhang}\ \emph {et~al.}(2011)\citenamefont {Zhang},
  \citenamefont {Lu},\ and\ \citenamefont {Xiang}}]{Zhang}%
  \BibitemOpen
  \bibfield  {author} {\bibinfo {author} {\bibfnamefont {G.~M.}\ \bibnamefont
  {Zhang}}, \bibinfo {author} {\bibfnamefont {Z.~Y.}\ \bibnamefont {Lu}}, \
  and\ \bibinfo {author} {\bibfnamefont {T.}~\bibnamefont {Xiang}},\ }\href
  {\doibase 10.1103/PhysRevB.84.052502} {\bibfield  {journal} {\bibinfo
  {journal} {Phys. Rev. B}\ }\textbf {\bibinfo {volume} {84}},\ \bibinfo
  {pages} {052502} (\bibinfo {year} {2011})}\BibitemShut {NoStop}%
\bibitem [{\citenamefont {Ye}\ \emph {et~al.}(2011)\citenamefont {Ye},
  \citenamefont {Chi}, \citenamefont {Bao}, \citenamefont {Wang}, \citenamefont
  {Ying}, \citenamefont {Chen}, \citenamefont {Wang}, \citenamefont {Dong},\
  and\ \citenamefont {Fang}}]{Ye}%
  \BibitemOpen
  \bibfield  {author} {\bibinfo {author} {\bibfnamefont {F.}~\bibnamefont
  {Ye}}, \bibinfo {author} {\bibfnamefont {S.}~\bibnamefont {Chi}}, \bibinfo
  {author} {\bibfnamefont {W.}~\bibnamefont {Bao}}, \bibinfo {author}
  {\bibfnamefont {X.~F.}\ \bibnamefont {Wang}}, \bibinfo {author}
  {\bibfnamefont {J.~J.}\ \bibnamefont {Ying}}, \bibinfo {author}
  {\bibfnamefont {X.~H.}\ \bibnamefont {Chen}}, \bibinfo {author}
  {\bibfnamefont {H.~D.}\ \bibnamefont {Wang}}, \bibinfo {author}
  {\bibfnamefont {C.~H.}\ \bibnamefont {Dong}}, \ and\ \bibinfo {author}
  {\bibfnamefont {M.}~\bibnamefont {Fang}},\ }\href {\doibase
  10.1103/PhysRevLett.107.137003} {\bibfield  {journal} {\bibinfo  {journal}
  {Phys. Rev. Lett.}\ }\textbf {\bibinfo {volume} {107}},\ \bibinfo {pages}
  {137003} (\bibinfo {year} {2011})}\BibitemShut {NoStop}%
\bibitem [{\citenamefont {Ryan}\ \emph {et~al.}(2011)\citenamefont {Ryan},
  \citenamefont {Rowan-Weetaluktuk}, \citenamefont {Cadogan}, \citenamefont
  {Hu}, \citenamefont {Straszheim}, \citenamefont {Bud'ko},\ and\ \citenamefont
  {Canfield}}]{Ryan}%
  \BibitemOpen
  \bibfield  {author} {\bibinfo {author} {\bibfnamefont {D.~H.}\ \bibnamefont
  {Ryan}}, \bibinfo {author} {\bibfnamefont {W.~N.}\ \bibnamefont
  {Rowan-Weetaluktuk}}, \bibinfo {author} {\bibfnamefont {J.~M.}\ \bibnamefont
  {Cadogan}}, \bibinfo {author} {\bibfnamefont {R.}~\bibnamefont {Hu}},
  \bibinfo {author} {\bibfnamefont {W.~E.}\ \bibnamefont {Straszheim}},
  \bibinfo {author} {\bibfnamefont {S.~L.}\ \bibnamefont {Bud'ko}}, \ and\
  \bibinfo {author} {\bibfnamefont {P.~C.}\ \bibnamefont {Canfield}},\ }\href
  {\doibase 10.1103/PhysRevB.83.104526} {\bibfield  {journal} {\bibinfo
  {journal} {Phys. Rev. B}\ }\textbf {\bibinfo {volume} {83}},\ \bibinfo
  {pages} {104526} (\bibinfo {year} {2011})}\BibitemShut {NoStop}%
\bibitem [{\citenamefont {Shermadini}\ \emph {et~al.}(2011)\citenamefont
  {Shermadini}, \citenamefont {Krzton-Maziopa}, \citenamefont {Bendele},
  \citenamefont {Khasanov}, \citenamefont {Luetkens}, \citenamefont {Conder},
  \citenamefont {Pomjakushina}, \citenamefont {Weyeneth}, \citenamefont
  {Pomjakushin}, \citenamefont {Bossen},\ and\ \citenamefont
  {Amato}}]{Shermadini}%
  \BibitemOpen
  \bibfield  {author} {\bibinfo {author} {\bibfnamefont {Z.}~\bibnamefont
  {Shermadini}}, \bibinfo {author} {\bibfnamefont {A.}~\bibnamefont
  {Krzton-Maziopa}}, \bibinfo {author} {\bibfnamefont {M.}~\bibnamefont
  {Bendele}}, \bibinfo {author} {\bibfnamefont {R.}~\bibnamefont {Khasanov}},
  \bibinfo {author} {\bibfnamefont {H.}~\bibnamefont {Luetkens}}, \bibinfo
  {author} {\bibfnamefont {K.}~\bibnamefont {Conder}}, \bibinfo {author}
  {\bibfnamefont {E.}~\bibnamefont {Pomjakushina}}, \bibinfo {author}
  {\bibfnamefont {S.}~\bibnamefont {Weyeneth}}, \bibinfo {author}
  {\bibfnamefont {V.}~\bibnamefont {Pomjakushin}}, \bibinfo {author}
  {\bibfnamefont {O.}~\bibnamefont {Bossen}}, \ and\ \bibinfo {author}
  {\bibfnamefont {A.}~\bibnamefont {Amato}},\ }\href {\doibase
  10.1103/PhysRevLett.106.117602} {\bibfield  {journal} {\bibinfo  {journal}
  {Phys. Rev. Lett.}\ }\textbf {\bibinfo {volume} {106}},\ \bibinfo {pages}
  {117602} (\bibinfo {year} {2011})}\BibitemShut {NoStop}%
\bibitem [{\citenamefont {Neilson}\ \emph {et~al.}(2012)\citenamefont
  {Neilson}, \citenamefont {Llobet}, \citenamefont {Stier}, \citenamefont {Wu},
  \citenamefont {Wen}, \citenamefont {Tao}, \citenamefont {Zhu}, \citenamefont
  {Tesanovic}, \citenamefont {Armitage},\ and\ \citenamefont
  {McQueen}}]{Neilson}%
  \BibitemOpen
  \bibfield  {author} {\bibinfo {author} {\bibfnamefont {J.~R.}\ \bibnamefont
  {Neilson}}, \bibinfo {author} {\bibfnamefont {A.}~\bibnamefont {Llobet}},
  \bibinfo {author} {\bibfnamefont {A.~V.}\ \bibnamefont {Stier}}, \bibinfo
  {author} {\bibfnamefont {L.}~\bibnamefont {Wu}}, \bibinfo {author}
  {\bibfnamefont {J.}~\bibnamefont {Wen}}, \bibinfo {author} {\bibfnamefont
  {J.}~\bibnamefont {Tao}}, \bibinfo {author} {\bibfnamefont {Y.}~\bibnamefont
  {Zhu}}, \bibinfo {author} {\bibfnamefont {Z.~B.}\ \bibnamefont {Tesanovic}},
  \bibinfo {author} {\bibfnamefont {N.~P.}\ \bibnamefont {Armitage}}, \ and\
  \bibinfo {author} {\bibfnamefont {T.~M.}\ \bibnamefont {McQueen}},\ }\href
  {\doibase 10.1103/PhysRevB.86.054512} {\bibfield  {journal} {\bibinfo
  {journal} {Phys. Rev. B}\ }\textbf {\bibinfo {volume} {86}},\ \bibinfo
  {pages} {054512} (\bibinfo {year} {2012})}\BibitemShut {NoStop}%
\bibitem [{\citenamefont {{Lei}}\ \emph {et~al.}(2012)\citenamefont {{Lei}},
  \citenamefont {{Wang}}, \citenamefont {{Ryu}}, \citenamefont {{Graf}},
  \citenamefont {{Warren}},\ and\ \citenamefont {{Petrovic}}}]{HLei}%
  \BibitemOpen
  \bibfield  {author} {\bibinfo {author} {\bibfnamefont {H.}~\bibnamefont
  {{Lei}}}, \bibinfo {author} {\bibfnamefont {K.}~\bibnamefont {{Wang}}},
  \bibinfo {author} {\bibfnamefont {H.}~\bibnamefont {{Ryu}}}, \bibinfo
  {author} {\bibfnamefont {D.}~\bibnamefont {{Graf}}}, \bibinfo {author}
  {\bibfnamefont {J.~B.}\ \bibnamefont {{Warren}}}, \ and\ \bibinfo {author}
  {\bibfnamefont {C.}~\bibnamefont {{Petrovic}}},\ }\href@noop {} {\bibfield
  {journal} {\bibinfo  {journal} {ArXiv e-prints}\ } (\bibinfo {year}
  {2012})},\ \Eprint {http://arxiv.org/abs/1211.1371} {arXiv:1211.1371
  [cond-mat.supr-con]} \BibitemShut {NoStop}%
\bibitem [{\citenamefont {Baroni}\ \emph {et~al.}(2001)\citenamefont {Baroni},
  \citenamefont {de~Gironcoli}, \citenamefont {Dal~Corso},\ and\ \citenamefont
  {Giannozzi}}]{DFPT}%
  \BibitemOpen
  \bibfield  {author} {\bibinfo {author} {\bibfnamefont {S.}~\bibnamefont
  {Baroni}}, \bibinfo {author} {\bibfnamefont {S.}~\bibnamefont
  {de~Gironcoli}}, \bibinfo {author} {\bibfnamefont {A.}~\bibnamefont
  {Dal~Corso}}, \ and\ \bibinfo {author} {\bibfnamefont {P.}~\bibnamefont
  {Giannozzi}},\ }\href {\doibase 10.1103/RevModPhys.73.515} {\bibfield
  {journal} {\bibinfo  {journal} {Rev. Mod. Phys.}\ }\textbf {\bibinfo {volume}
  {73}},\ \bibinfo {pages} {515} (\bibinfo {year} {2001})}\BibitemShut
  {NoStop}%
\bibitem [{\citenamefont {Giannozzi}\ \emph {et~al.}(2009)\citenamefont
  {Giannozzi}, \citenamefont {Baroni}, \citenamefont {Bonini}, \citenamefont
  {Calandra}, \citenamefont {Car}, \citenamefont {Cavazzoni}, \citenamefont
  {Ceresoli}, \citenamefont {Chiarotti}, \citenamefont {Cococcioni},
  \citenamefont {Dabo}, \citenamefont {Corso}, \citenamefont {de~Gironcoli},
  \citenamefont {Fabris}, \citenamefont {Fratesi}, \citenamefont {Gebauer},
  \citenamefont {Gerstmann}, \citenamefont {Gougoussis}, \citenamefont
  {Kokalj}, \citenamefont {Lazzeri}, \citenamefont {Martin-Samos},
  \citenamefont {Marzari}, \citenamefont {Mauri}, \citenamefont {Mazzarello},
  \citenamefont {Paolini}, \citenamefont {Pasquarello}, \citenamefont
  {Paulatto}, \citenamefont {Sbraccia}, \citenamefont {Scandolo}, \citenamefont
  {Sclauzero}, \citenamefont {Seitsonen}, \citenamefont {Smogunov},
  \citenamefont {Umari},\ and\ \citenamefont {Wentzcovitch}}]{QE}%
  \BibitemOpen
  \bibfield  {author} {\bibinfo {author} {\bibfnamefont {P.}~\bibnamefont
  {Giannozzi}}, \bibinfo {author} {\bibfnamefont {S.}~\bibnamefont {Baroni}},
  \bibinfo {author} {\bibfnamefont {N.}~\bibnamefont {Bonini}}, \bibinfo
  {author} {\bibfnamefont {M.}~\bibnamefont {Calandra}}, \bibinfo {author}
  {\bibfnamefont {R.}~\bibnamefont {Car}}, \bibinfo {author} {\bibfnamefont
  {C.}~\bibnamefont {Cavazzoni}}, \bibinfo {author} {\bibfnamefont
  {D.}~\bibnamefont {Ceresoli}}, \bibinfo {author} {\bibfnamefont {G.~L.}\
  \bibnamefont {Chiarotti}}, \bibinfo {author} {\bibfnamefont {M.}~\bibnamefont
  {Cococcioni}}, \bibinfo {author} {\bibfnamefont {I.}~\bibnamefont {Dabo}},
  \bibinfo {author} {\bibfnamefont {A.~D.}\ \bibnamefont {Corso}}, \bibinfo
  {author} {\bibfnamefont {S.}~\bibnamefont {de~Gironcoli}}, \bibinfo {author}
  {\bibfnamefont {S.}~\bibnamefont {Fabris}}, \bibinfo {author} {\bibfnamefont
  {G.}~\bibnamefont {Fratesi}}, \bibinfo {author} {\bibfnamefont
  {R.}~\bibnamefont {Gebauer}}, \bibinfo {author} {\bibfnamefont
  {U.}~\bibnamefont {Gerstmann}}, \bibinfo {author} {\bibfnamefont
  {C.}~\bibnamefont {Gougoussis}}, \bibinfo {author} {\bibfnamefont
  {A.}~\bibnamefont {Kokalj}}, \bibinfo {author} {\bibfnamefont
  {M.}~\bibnamefont {Lazzeri}}, \bibinfo {author} {\bibfnamefont
  {L.}~\bibnamefont {Martin-Samos}}, \bibinfo {author} {\bibfnamefont
  {N.}~\bibnamefont {Marzari}}, \bibinfo {author} {\bibfnamefont
  {F.}~\bibnamefont {Mauri}}, \bibinfo {author} {\bibfnamefont
  {R.}~\bibnamefont {Mazzarello}}, \bibinfo {author} {\bibfnamefont
  {S.}~\bibnamefont {Paolini}}, \bibinfo {author} {\bibfnamefont
  {A.}~\bibnamefont {Pasquarello}}, \bibinfo {author} {\bibfnamefont
  {L.}~\bibnamefont {Paulatto}}, \bibinfo {author} {\bibfnamefont
  {C.}~\bibnamefont {Sbraccia}}, \bibinfo {author} {\bibfnamefont
  {S.}~\bibnamefont {Scandolo}}, \bibinfo {author} {\bibfnamefont
  {G.}~\bibnamefont {Sclauzero}}, \bibinfo {author} {\bibfnamefont {A.~P.}\
  \bibnamefont {Seitsonen}}, \bibinfo {author} {\bibfnamefont {A.}~\bibnamefont
  {Smogunov}}, \bibinfo {author} {\bibfnamefont {P.}~\bibnamefont {Umari}}, \
  and\ \bibinfo {author} {\bibfnamefont {R.~M.}\ \bibnamefont {Wentzcovitch}},\
  }\href {http://stacks.iop.org/0953-8984/21/i=39/a=395502} {\bibfield
  {journal} {\bibinfo  {journal} {Journal of Physics: Condensed Matter}\
  }\textbf {\bibinfo {volume} {21}},\ \bibinfo {pages} {395502} (\bibinfo
  {year} {2009})}\BibitemShut {NoStop}%
\bibitem [{\citenamefont {Lazarevi\ifmmode~\acute{c}\else \'{c}\fi{}}\ \emph
  {et~al.}(2012)\citenamefont {Lazarevi\ifmmode~\acute{c}\else \'{c}\fi{}},
  \citenamefont {Abeykoon}, \citenamefont {Stephens}, \citenamefont {Lei},
  \citenamefont {Bozin}, \citenamefont {Petrovic},\ and\ \citenamefont
  {Popovi\ifmmode~\acute{c}\else \'{c}\fi{}}}]{Lazarevic1}%
  \BibitemOpen
  \bibfield  {author} {\bibinfo {author} {\bibfnamefont {N.}~\bibnamefont
  {Lazarevi\ifmmode~\acute{c}\else \'{c}\fi{}}}, \bibinfo {author}
  {\bibfnamefont {M.}~\bibnamefont {Abeykoon}}, \bibinfo {author}
  {\bibfnamefont {P.~W.}\ \bibnamefont {Stephens}}, \bibinfo {author}
  {\bibfnamefont {H.}~\bibnamefont {Lei}}, \bibinfo {author} {\bibfnamefont
  {E.~S.}\ \bibnamefont {Bozin}}, \bibinfo {author} {\bibfnamefont
  {C.}~\bibnamefont {Petrovic}}, \ and\ \bibinfo {author} {\bibfnamefont
  {Z.~V.}\ \bibnamefont {Popovi\ifmmode~\acute{c}\else \'{c}\fi{}}},\ }\href
  {\doibase 10.1103/PhysRevB.86.054503} {\bibfield  {journal} {\bibinfo
  {journal} {Phys. Rev. B}\ }\textbf {\bibinfo {volume} {86}},\ \bibinfo
  {pages} {054503} (\bibinfo {year} {2012})}\BibitemShut {NoStop}%
\bibitem [{\citenamefont {Lazarevi\ifmmode~\acute{c}\else \'{c}\fi{}}\ \emph
  {et~al.}(2011)\citenamefont {Lazarevi\ifmmode~\acute{c}\else \'{c}\fi{}},
  \citenamefont {Lei}, \citenamefont {Petrovic},\ and\ \citenamefont
  {Popovi\ifmmode~\acute{c}\else \'{c}\fi{}}}]{Lazarevic2}%
  \BibitemOpen
  \bibfield  {author} {\bibinfo {author} {\bibfnamefont {N.}~\bibnamefont
  {Lazarevi\ifmmode~\acute{c}\else \'{c}\fi{}}}, \bibinfo {author}
  {\bibfnamefont {H.}~\bibnamefont {Lei}}, \bibinfo {author} {\bibfnamefont
  {C.}~\bibnamefont {Petrovic}}, \ and\ \bibinfo {author} {\bibfnamefont
  {Z.~V.}\ \bibnamefont {Popovi\ifmmode~\acute{c}\else \'{c}\fi{}}},\ }\href
  {\doibase 10.1103/PhysRevB.84.214305} {\bibfield  {journal} {\bibinfo
  {journal} {Phys. Rev. B}\ }\textbf {\bibinfo {volume} {84}},\ \bibinfo
  {pages} {214305} (\bibinfo {year} {2011})}\BibitemShut {NoStop}%
\bibitem [{\citenamefont {Balkanski}\ \emph {et~al.}(1983)\citenamefont
  {Balkanski}, \citenamefont {Wallis},\ and\ \citenamefont {Haro}}]{anh1}%
  \BibitemOpen
  \bibfield  {author} {\bibinfo {author} {\bibfnamefont {M.}~\bibnamefont
  {Balkanski}}, \bibinfo {author} {\bibfnamefont {R.~F.}\ \bibnamefont
  {Wallis}}, \ and\ \bibinfo {author} {\bibfnamefont {E.}~\bibnamefont
  {Haro}},\ }\href {\doibase 10.1103/PhysRevB.28.1928} {\bibfield  {journal}
  {\bibinfo  {journal} {Phys. Rev. B}\ }\textbf {\bibinfo {volume} {28}},\
  \bibinfo {pages} {1928} (\bibinfo {year} {1983})}\BibitemShut {NoStop}%
\bibitem [{\citenamefont {Lazarevi\ifmmode~\acute{c}\else \'{c}\fi{}}\ \emph
  {et~al.}(2010)\citenamefont {Lazarevi\ifmmode~\acute{c}\else \'{c}\fi{}},
  \citenamefont {Popovi\ifmmode~\acute{c}\else \'{c}\fi{}}, \citenamefont
  {Hu},\ and\ \citenamefont {Petrovic}}]{LazarevicFano}%
  \BibitemOpen
  \bibfield  {author} {\bibinfo {author} {\bibfnamefont {N.}~\bibnamefont
  {Lazarevi\ifmmode~\acute{c}\else \'{c}\fi{}}}, \bibinfo {author}
  {\bibfnamefont {Z.~V.}\ \bibnamefont {Popovi\ifmmode~\acute{c}\else
  \'{c}\fi{}}}, \bibinfo {author} {\bibfnamefont {R.}~\bibnamefont {Hu}}, \
  and\ \bibinfo {author} {\bibfnamefont {C.}~\bibnamefont {Petrovic}},\ }\href
  {\doibase 10.1103/PhysRevB.81.144302} {\bibfield  {journal} {\bibinfo
  {journal} {Phys. Rev. B}\ }\textbf {\bibinfo {volume} {81}},\ \bibinfo
  {pages} {144302} (\bibinfo {year} {2010})}\BibitemShut {NoStop}%
\end{thebibliography}
%

\end{document}